\documentclass[conference]{IEEEtran}
\IEEEoverridecommandlockouts

\usepackage{cite}
\usepackage{amsmath,amssymb,amsfonts}
\usepackage{algorithmic}
\usepackage{algorithm}
\usepackage{graphicx}
\usepackage{subcaption}
\usepackage{textcomp}
\usepackage{xcolor}
\usepackage{booktabs}
\usepackage[hidelinks]{hyperref}

\def\BibTeX{{\rm B\kern-.05em{\sc i\kern-.025em b}\kern-.08em
T\kern-.1667em\lower.7ex\hbox{E}\kern-.125emX}}

\begin{document}

\title{Hardware-in-the-Loop Evaluation of Game-Theoretic Autonomous Driving}

\author{
\IEEEauthorblockN{Aakanksha Kataria\textsuperscript{1}, Huiwen Yan\textsuperscript{2}, and Mushuang Liu\textsuperscript{2}}
\IEEEauthorblockA{
\textit{Virginia Tech, Blacksburg, VA, USA}\\
\textsuperscript{1}\textit{Department of Electrical and Computer Engineering}\\
\textsuperscript{2}\textit{Department of Mechanical Engineering}
}
}

\maketitle

\begin{abstract}
This paper evaluates Nash- and Stackelberg-based decision-making controllers for autonomous intersection crossing using a three-stage evaluation pipeline culminating in physical Quanser QCar 2 experiments with hardware-in-the-loop (HIL) execution. The controllers are implemented in MATLAB/Simulink, deployed through Quanser Real-Time Control (QUARC) software, and executed on the onboard NVIDIA Jetson AGX Orin processor. The evaluation includes MATLAB numerical simulation, qualitative validation in Quanser Interactive Labs (QLabs), and physical QCar 2 experiments. The experiments consider symmetric and asymmetric intersection approaches, leader-follower interactions, conflicting Stackelberg role assignments, and non-cooperative obstacle-vehicle behaviors. 
The results characterize the effects of hierarchy assignment, obstacle-vehicle behavior, and physical implementation on the considered game-theoretic autonomous driving controllers. Comparison between software simulations and hardware experiments further highlights the importance of accounting for sensing and state-estimation uncertainty when translating game-theoretic controllers from simulation to physical systems. A video demonstration of the QLabs simulations and physical QCar~2 hardware experiments is available at \textit{\textbf{\url{https://youtu.be/gkV6lz0twRk}}}.
\end{abstract}


\section{Introduction}
\label{sec:introduction}

Autonomous vehicles are expected to make safe and efficient decisions when interacting with other road users. Intersection crossing is a representative interaction scenario because vehicles may have conflicting paths and incomplete information regarding other vehicles' intentions. Game-theoretic approaches have been used to model such interactions by representing each vehicle as a player with an individual driving objective \cite{namazi2019,muhlethaler2026,tian2019,li2019}.

Nash and Stackelberg games are two commonly used formulations for autonomous-driving decision-making. In a Nash game, players make decisions simultaneously, and each player selects a best response to the strategies of the other players  \cite{liu2022_potential}. In a Stackelberg game, one or more players are assigned as leaders, and the remaining players respond as followers \cite{bateman2023,hang2020}. The two formulations differ in their assumptions regarding road priority.

For Nash-based decision-making, potential-game formulations are particularly relevant to our work because of their theoretical guarantees. Sufficient conditions for constructing potential games and the corresponding formulations for autonomous driving are developed in \cite{liu2022_potential}, guaranteeing the existence of a pure-strategy Nash equilibrium. Related studies have explored predictor-corrector potential games under incomplete information \cite{liu2022_predictor}, game-projection methods \cite{liu2023_projection}, mixed-integer potential games \cite{fabiani2020}, and Markov potential games for multi-agent learning \cite{yan2026}. Despite these theoretical advances, the proposed methods have been evaluated primarily through numerical simulations.

Stackelberg games provide a formulation for scenarios with an available leader-follower relationship, but the resulting behavior depends on the assigned hierarchy and assumed follower behavior \cite{bateman2023,hang2020,geary2020}. Simple comparative studies between Nash and Stackelberg games in intersection-crossing scenarios were evaluated using safety, travel efficiency, and computational time as performance metrics \cite{bateman2023}. Their results showed that multi-player Nash games may provide improved robustness relative to pairwise games when traffic-rule violations occur, although the computational cost increases with the number of players.

Computational cost is an important and practical performance metric for game-theoretic decision-making in autonomous driving. The computational demands of Nash-based decision-making can increase substantially with the number of players and available strategies \cite{bateman2023}. Methods based on time-distributed iterations and decomposed decision-making have been investigated to reduce online computational requirements \cite{liu2026_time,suriyarachchi2022}. However, controller feasibility also depends on the hardware: the processor, sensors, actuation systems, and execution timing of the platform being used \cite{liu2020_computing}.

Software simulation alone does not fully represent physical vehicle operations. Differences between simulated and physical systems can result from vehicle dynamics, sensing errors, actuator delays, communication delays, and environmental conditions \cite{stocco2021,hu2023,lambertenghi2025}. These differences can affect a game-theoretic controller because its selected action depends on the estimated state and predicted actions of the interacting vehicles. Hardware evaluation is therefore critical to determine whether interaction behaviors observed in software simulations are preserved after deployment.

This paper evaluates Nash and Stackelberg intersection controllers using a three-stage pipeline. The controllers are first evaluated in MATLAB, then QLabs, and finally deployed on a Quanser QCar 2 platform \cite{quanser_qcar2}. The contributions include:
\begin{enumerate}
    \item Nash and Stackelberg controllers for intersection scenarios are implemented in MATLAB/Simulink and deployed through QUARC on a Quanser QCar 2 equipped with an NVIDIA Jetson AGX Orin processor.
    \item A three-stage evaluation pipeline comparing MATLAB simulation, QLabs validation, and hardware experiments is used to assess controller behavior from numerical simulation to physical vehicle operation.
    \item QLabs is used to verify controller–vehicle integration, including control interfaces, coordinate conventions, command scaling, and compatibility with vehicle mechanics, before physical deployment.
    \item The effects of symmetric interactions, consistent and conflicting Stackelberg role assignments, and non-cooperative obstacle-vehicle behaviors are reported.
\end{enumerate}

\section{Game-Theoretic Controller Design}
\label{sec:methodology}

This section presents the receding-horizon decision-making formulation. At each decision step, the controller predicts trajectories for a finite set of candidate actions and selects an action based on either a Nash or a Stackelberg game formulation.

\subsection{Vehicle Dynamics and Cost Functions}\label{subsec:Designs}
\begin{figure}[!t]
\centering
\includegraphics[width=0.6\linewidth]{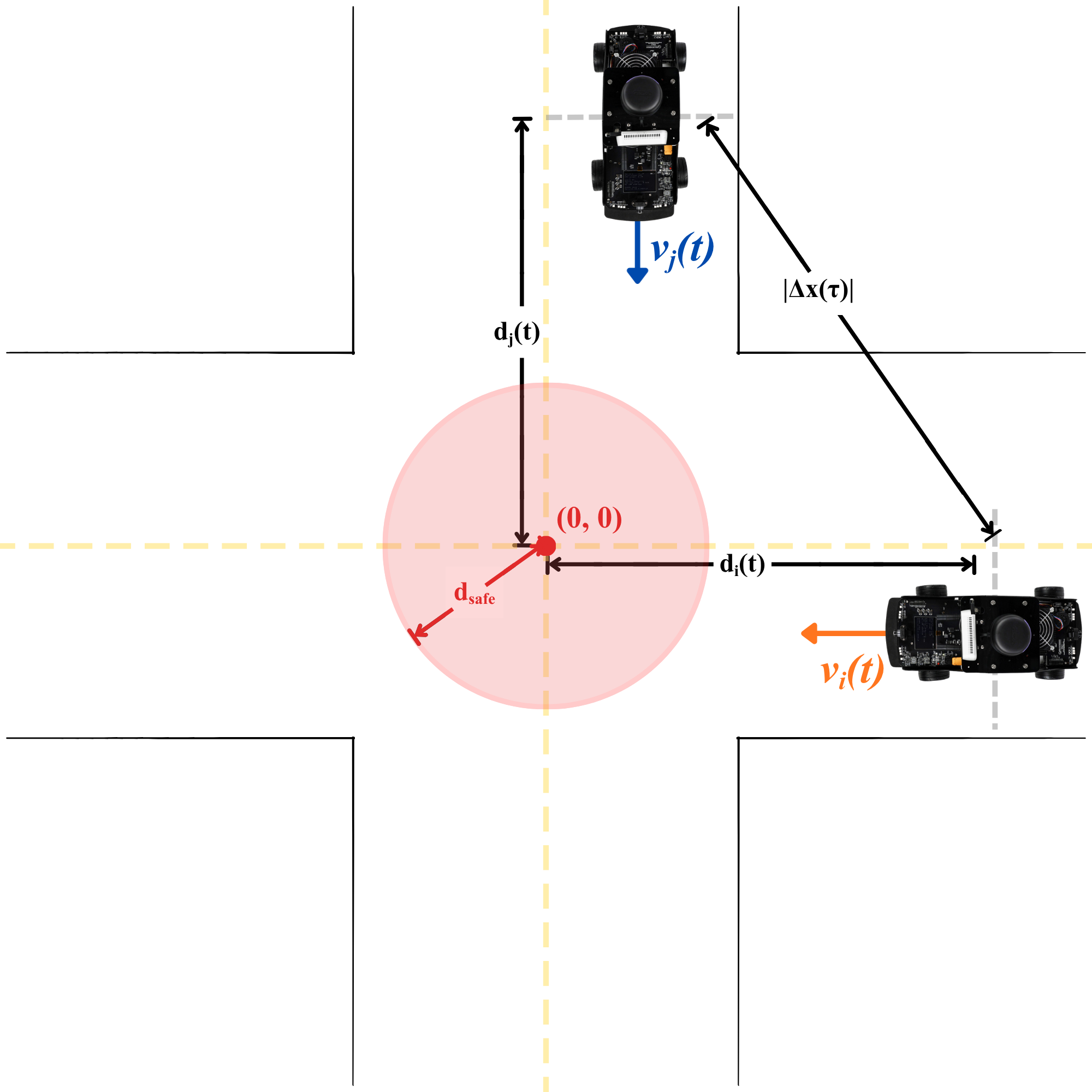}
\caption{Geometry of the considered intersection. Vehicle $i$ and Vehicle $j$ travel along fixed straight paths toward the intersection center $(0,0)$. 
}
\label{fig:intersection_geometry}
\end{figure}

Consider a set of vehicles $\mathcal{N}=\{1,\ldots,N\}$, where $N$ is the number of vehicles participating in the game. The decision-making controller uses a discrete-time longitudinal kinematic model. Let $d_i(t)$ denote the remaining distance from vehicle $i$ to the center of the intersection (Fig. \ref{fig:intersection_geometry}), $v_i(t)$ denote the longitudinal speed, and $u_i(t)$ denote the longitudinal acceleration input. The dynamics used for trajectory prediction are
\begin{equation}
v_i(t+1)=\max\left(v_i(t)+u_i(t)\Delta t,0\right),
\label{eq:velocity_dynamics}
\end{equation}
\begin{equation}
d_i(t+1)=d_i(t)-v_i(t+1)\Delta t,
\label{eq:distance_dynamics}
\end{equation}

where $\Delta t=0.1$ s is the sampling period. The state of vehicle $i$ is defined as $x_i(t)=[d_i(t),v_i(t)]^\top$. Each candidate acceleration is held constant over the prediction horizon of $T=15$ steps, corresponding to $1.5$ s.

For the intersection scenarios considered in our work, we let $\mathcal{U}_i$ denote the finite set of candidate longitudinal acceleration inputs available to vehicle $i$. At each decision step $t$, each candidate input $u_i(t)\in\mathcal{U}_i$ is held constant over the prediction horizon when computing the predicted trajectory. The joint action is
$\mathbf{u}(t)=\bigl(u_1(t),\ldots,u_N(t)\bigr)\in\mathcal{U}$
, where $\mathcal{U}=\mathcal{U}_1\times\cdots\times\mathcal{U}_N$ is the joint action space.

Each vehicle seeks to minimize its own cost according to the adopted game formulation. The cost comprises speed-tracking, control-effort, and collision-avoidance terms:
\begin{equation}
\begin{aligned}
J_i^t(\mathbf{u}(t))
={}&\theta_1 J_{\mathrm{speed},i}^t(\mathbf{u}(t))\\
&+\theta_2 J_{\mathrm{effort},i}^t(\mathbf{u}(t))
+\theta_3 J_{\mathrm{collision},i}^t(\mathbf{u}(t)),
\end{aligned}
\label{eq}
\end{equation}
where $\theta_1,\theta_2,\theta_3> 0$ are weighting parameters.

The speed-tracking term is
\begin{equation}
J_{\mathrm{speed},i}^t(\mathbf{u}(t)) =
\sum_{\tau=t}^{t+T-1}
\left(
\frac{v_i(\tau)-v_{\mathrm{desired}}}
{v_{\mathrm{desired}}}
\right)^2,
\label{eq:speed_cost}
\end{equation}
where $v_{\mathrm{desired}}$ is vehicle $i$'s desired speed.

The control-effort term is
\begin{equation}
J_{\mathrm{effort},i}(\mathbf{u}(t)) = \sum_{\tau=t}^{t+T-1}a_i(t)^2,
\label{eq:effort_cost}
\end{equation}
where $a_i$ is the longitudinal acceleration of the vehicle.

The collision-avoidance term penalizes predicted proximity between vehicle $i$ and its obstacle vehicle:
\begin{equation}
J_{\mathrm{collision},i}^t(\mathbf{u}(t)) =
\sum_{\tau=t}^{t+T-1}
\left[
\tanh\left(
\beta(d_{\mathrm{safe}}-|\Delta x(\tau)|)
\right)+1
\right],
\label{eq:collision_cost}
\end{equation}
where $d_{\mathrm{safe}}$ is the selected safety distance, $\beta$ is used to control the transition of the penalty function, and $|\Delta x|$ denotes the predicted Euclidean distance between the centers of mass of the two vehicles.

We represent the game at time $t$ as
\begin{equation}
\mathcal{G}^t=\left\{\mathcal{N},\mathcal{U},\{J_i^t\}_{i\in\mathcal{N}}\right\}.
\label{eq:game_definition}
\end{equation}
We denote Nash and Stackelberg formulations as $\mathcal{G}^{N,t}$ and $\mathcal{G}^{S,t}$, respectively.
\subsection{Nash Equilibrium Controller}

The Nash controller is used when the vehicles are treated symmetrically. We formulate the interaction as a finite potential game. According to Theorem 3 in \cite{liu2022_potential}, the cost function designs in \ref{subsec:Designs} yield a finite potential game with the following potential function:
\begin{equation}
\begin{aligned}
&F^t(\mathbf{u}(t)) =
\sum_{i\in\mathcal{N}} \Big(J_{\mathrm{speed},i}^t(\mathbf{u}(t))+J_{\mathrm{effort},i}(\mathbf{u}(t))\Big)\\
&\quad+
\sum_{i\in\mathcal{N}}
\sum_{\substack{j\in\mathcal{N}\\j<i}}
J_{\mathrm{collision},i}^t(\mathbf{u}(t)).
\label{eq:potential}
\end{aligned}
\end{equation}
For the finite action set considered in this work, the controller evaluates the candidate action profiles and selects
\begin{equation}
\mathbf{u}^*(t) \in \arg\min_{\mathbf{u}(t)\in\mathcal{U}} F^t(\mathbf{u}(t)).
\label{eq:nash_solution}
\end{equation}
A global minimizer of the exact potential function is a pure-strategy Nash equilibrium under the potential-game formulation \cite{liu2022_potential}. The detailed steps are shown in Algorithm \ref{alg:potential}.
\begin{algorithm}
\caption{Potential Function Optimization}
\label{alg:potential}
\begin{algorithmic}[1]
\STATE \textbf{Input:} $\mathcal{G}^{N,t}$
\STATE \textbf{Output:} $u_{\mathrm{ego}}^*(t)$, optimal ego acceleration
\STATE Initialize $F_{\min}\leftarrow\infty$
\FOR{each $u_{\mathrm{ego}}\in\mathcal{U}_{\mathrm{ego}}$}
\FOR{each $u_{\mathrm{obs}}\in\mathcal{U}_{\mathrm{obs}}$}
\STATE Predict vehicle trajectories over horizon $T$
\STATE Compute $F^t$
\IF{$F^t<F_{\min}$}
\STATE $F_{\min}\leftarrow F^t$
\STATE $u_{\mathrm{ego}}^*(t)\leftarrow u_{\mathrm{ego}}$
\ENDIF
\ENDFOR
\ENDFOR
\RETURN $u_{\mathrm{ego}}^*(t)$
\end{algorithmic}
\end{algorithm}
\subsection{Stackelberg Equilibrium Controller}
The Stackelberg controller is used when a leader-follower order is assigned. The leader selects an action while anticipating the follower's best response. In this work, the role
assignment is specified by the binary input
$\texttt{ego\_is\_leader}\in\{0,1\}$. When
$\texttt{ego\_is\_leader}=1$, the ego vehicle is the leader and the
obstacle vehicle is the follower; otherwise, the roles are reversed.

Let $\mathcal{U}_L$ and $\mathcal{U}_F$ denote the candidate action sets of the leader and follower, respectively. For each candidate leader action $u_L\in\mathcal{U}_L$, let $u_F^{*u_L}$ denote the follower action that minimizes the follower cost for the given leader action. The follower’s best response is
\begin{equation}
\mathrm{BR}(u_L)=
\arg\min_{u_F\in\mathcal{U}_F}
J_F^t(u_L,u_F).
\label{eq:best_response}
\end{equation}
Let $u_F^{*u_L}\in\mathrm{BR}(u_L)$ denote the follower action selected from the best-response set for the candidate leader action $u_L$.

For a strong Stackelberg equilibrium, if the follower has multiple best responses, the one minimizing the leader's cost is selected. Therefore, the leader selects according to
\begin{equation}
u_L^*(t)=
\arg\min_{u_L\in\mathcal{U}_L}
J_L^t(u_L,u_F^{*u_L}).
\label{eq:stackelberg_solution}
\end{equation}

The ego-vehicle action returned by the controller depends on its assigned role. If the ego vehicle is the leader, the controller returns $u_L^*(t)$. Otherwise, it returns the follower best response $u_F^*(t)$ corresponding to the selected leader action. The detailed steps are shown in Algorithm \ref{alg:stackelberg}.

\begin{algorithm}
\caption{Strong Stackelberg Equilibrium for Two Players}
\label{alg:stackelberg}
\begin{algorithmic}[1]
\STATE \textbf{Input:} $\mathcal{G}^{S,t}$, $\texttt{ego\_is\_leader}$
\STATE \textbf{Output:} $u_{\mathrm{ego}}^*(t)$, optimal ego acceleration
\STATE Initialize $J_{L,\min}\leftarrow\infty$
\FOR{each $u_L\in\mathcal{U}_L$}
    \STATE $J_{F,\min}\leftarrow\infty$
    \FOR{each $u_F\in\mathcal{U}_F$}
        \STATE Predict vehicle trajectories over horizon $T$
        \STATE Compute the follower cost $J_F$
        \IF{$J_F<J_{F,\min}$}
            \STATE $J_{F,\min}\leftarrow J_F$
            \STATE $u_F^{*u_L}\leftarrow u_F$
        \ENDIF
    \ENDFOR
    \STATE Predict trajectories using $u_L$ and $u_F^{*u_L}$
    \STATE Compute the leader cost $J_L$
    \IF{$J_L<J_{L,\min}$}
        \STATE $J_{L,\min}\leftarrow J_L$
        \STATE $u_L^*(t)\leftarrow u_L$
        \STATE $u_F^*(t)\leftarrow u_F^{*u_L}$
    \ENDIF
\ENDFOR
\IF{$\texttt{ego\_is\_leader}=1$}
    \STATE $u_{\mathrm{ego}}^*(t)\leftarrow u_L^*(t)$
\ELSE
    \STATE $u_{\mathrm{ego}}^*(t)\leftarrow u_F^*(t)$
\ENDIF
\RETURN $u_{\mathrm{ego}}^*(t)$
\end{algorithmic}
\end{algorithm}

\section{System Architecture and Evaluation Pipeline}
\label{sec:architecture}
This section describes the proposed system architecture and evaluation pipeline. It first defines the intersection scenario and vehicle model, then presents the game-theoretic decision process, and finally outlines the simulation setup and metrics used to evaluate safety and performance.

\subsection{Hardware-in-the-Loop Platform}

\begin{figure}[!t]
\centering
\includegraphics[width=1\linewidth]{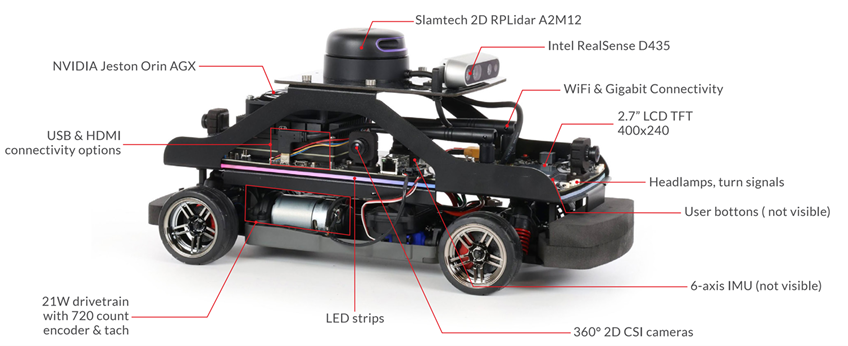}
\caption{Hardware architecture of the Quanser QCar 2 platform.} 
\label{fig:qcar2_hardware}
\end{figure}

The controllers are deployed on a Quanser QCar 2 \cite{quanser_qcar2}, a one-tenth-scale autonomous vehicle platform equipped with an NVIDIA Jetson AGX Orin onboard computer, an Intel RealSense D435 RGB-D camera, a two-dimensional light detection and ranging (LiDAR) sensor, wheel encoders, and an inertial measurement unit (IMU) (see Figure \ref{fig:qcar2_hardware}). 

For the hardware-in-the-loop experiments, the controllers are implemented in MATLAB/Simulink, deployed through Quanser Real-Time Control (QUARC) software, and executed on the onboard computer. The closed-loop system incorporates onboard sensing, state estimation, game-theoretic decision-making, command generation, drivetrain actuation, and physical vehicle motion.

The deployed control loop operates at 100 Hz. At each control step, the system updates the vehicle state estimate, evaluates the game-theoretic decision rule, and applies the resulting control command. The reported execution times measure the game-theoretic controller's computation time during steady-state operation, excluding one-time initialization.

\subsection{Three-Stage Evaluation Pipeline}

The evaluation consists of three stages:
\begin{figure}[!t]
\centering
\includegraphics[width=0.8\linewidth]{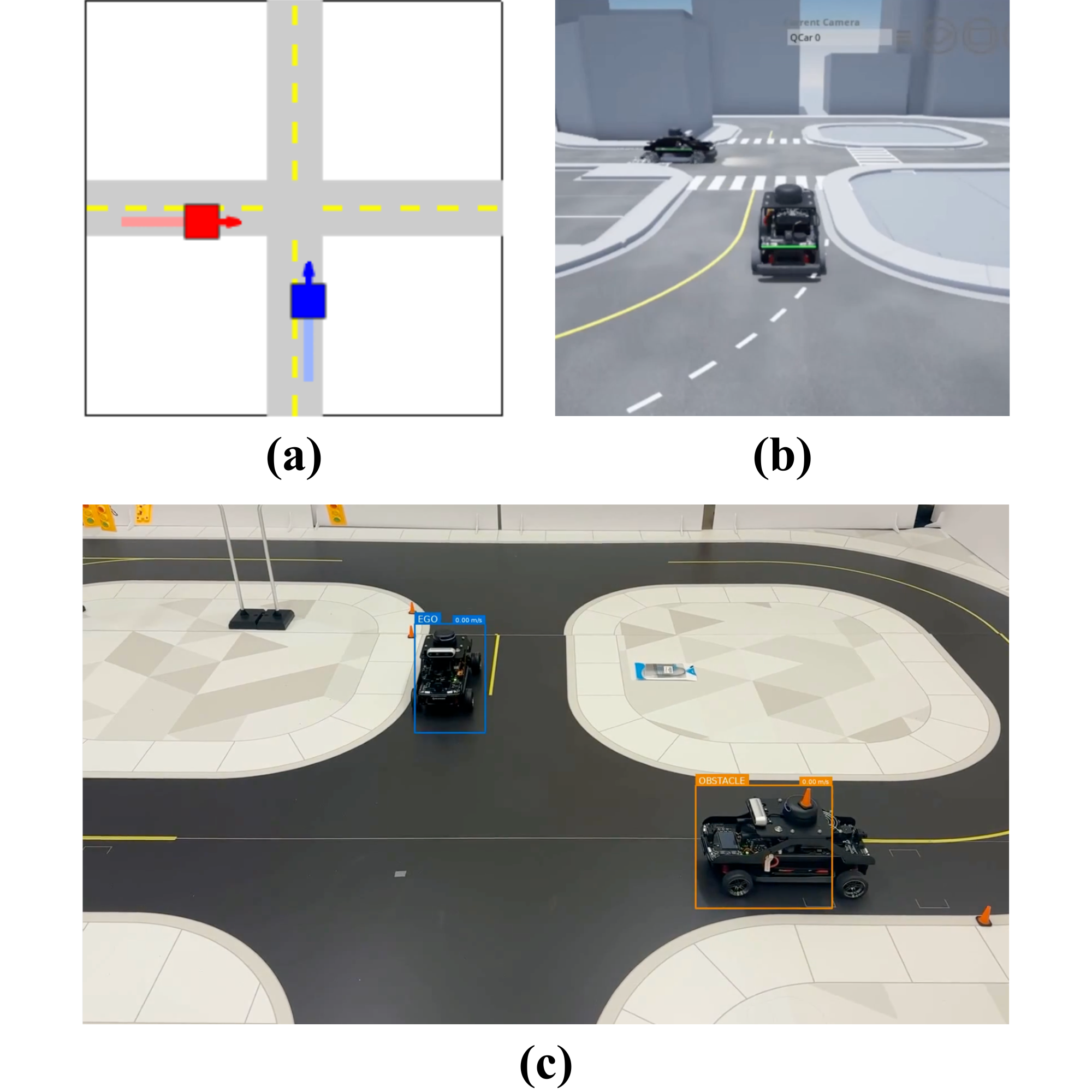}
\caption{Evaluation environments: (a) MATLAB simulation, (b) QLabs virtual validation, and (c) physical hardware testbed evaluation using QCar 2.}
\label{fig:three_stage_pipeline}
\end{figure}

\begin{enumerate}
\item \textbf{MATLAB simulation:} A kinematic vehicle model is used to evaluate the game formulations and candidate trajectories.
\item \textbf{Quanser Interactive Labs (QLabs) validation:} The MATLAB/Simulink controllers are integrated with the QLabs vehicle model to verify control interfaces, coordinate conventions, command scaling, and compatibility with the modeled vehicle mechanics before physical deployment.
\item \textbf{QCar 2 hardware experiments:} The controllers are executed on the physical QCar 2 platform to evaluate their performance under onboard sensing, physical actuation, and embedded computation.
\end{enumerate}

The quantitative results reported in this paper are obtained from MATLAB simulations and QCar 2 hardware experiments. The QLabs stage provides qualitative validation and helps identify implementation and interface errors during the transition from numerical simulation to physical deployment.

\section{Experimental Setup}
\label{sec:experimental_setup}
In this section, we design the test scenarios used to evaluate the proposed Nash and Stackelberg controllers. All scenarios consider an ego vehicle and an obstacle vehicle with intersecting paths. The test cases include Nash interactions, Stackelberg leader-follower interactions, conflicting Stackelberg role assignments, and obstacle vehicles with prescribed non-cooperative behaviors.
\subsection{MATLAB Numerical Simulation}
The MATLAB simulation uses a kinematic vehicle model. Each scenario is repeated for 100 trials, and the reported metrics are averaged across the trials. The MATLAB stage evaluates the game-theoretic decision logic without sensing errors or uncertainties.
The following test cases are considered:
\begin{itemize}
\item \textbf{Test 1: Nash interaction}
\begin{itemize}
\item \textit{Run 1A: Symmetric arrival.} Both vehicles begin at $0.0$ m/s. This case evaluates the Nash controller under symmetric initial conditions.
\item \textit{Run 1B: Asymmetric arrival.} The ego vehicle begins at $0.4$ m/s and the obstacle vehicle begins at $0.0$ m/s. This case evaluates whether an initial speed difference produces a passing order.
\end{itemize}

\item \textbf{Test 2: Stackelberg leader-follower interaction}
\begin{itemize}
\item \textit{Run 2A: Symmetric leader-follower case.} Both vehicles begin at $0.0$ m/s. The ego vehicle is assigned as the Stackelberg leader and the obstacle vehicle is assigned as the Stackelberg follower.
\item \textit{Run 2B: Asymmetric leader-follower case.} The Stackelberg leader begins at $0.0$ m/s and the follower begins at $0.3$ m/s. This case evaluates whether the collision cost causes the leader to yield when the follower has a proceeding advantage.
\end{itemize}

\item \textbf{Test 3: Stackelberg hierarchy conflict}
\begin{itemize}
\item \textit{Run 3: Leader-leader conflict.} Both vehicles begin at $0.0$ m/s and are assigned leader roles. This case evaluates the result of incompatible hierarchy assignments.
\end{itemize}

\item \textbf{Test 4: Robustness cases}
\begin{itemize}
\item \textit{Runs 4A and 4B: Constant-speed obstacle.} The ego vehicle begins at $0.0$ m/s. The obstacle vehicle maintains a prescribed speed of $0.4$ m/s and does not respond to the ego vehicle. Runs 4A and 4B evaluate the Nash and Stackelberg-leader configurations, respectively.
\item \textit{Runs 4C and 4D: Unexpected-braking obstacle.} The obstacle vehicle initially approaches the intersection at a higher speed and subsequently brakes. These runs evaluate whether the ego vehicle updates its selected action after the obstacle-vehicle behavior changes.
\end{itemize}
\end{itemize}

\subsection{QLabs Simulation}
The MATLAB/Simulink controller is connected to the QLabs vehicle model before deployment on the physical QCar 2 platform.
The QLabs simulations include the following test cases:
\begin{itemize}
\item \textbf{Nash interaction:} The symmetric case is used to verify that the integrated controller produces the expected collision-avoidance and yielding behavior in the QLabs vehicle environment.
\item \textbf{Stackelberg leader versus Nash obstacle intersection:} The ego vehicle executes the Stackelberg controller with the leader role, while the obstacle vehicle executes the Nash controller. This case is used to verify that the implemented controller pairing produces the intended passing order in the QLabs vehicle environment.
\end{itemize}

A QLabs test is considered successful when the deployed Simulink controller produces the expected qualitative vehicle behavior without interface errors, command-sign errors, inconsistent coordinate transformations, or unintended vehicle-motion behavior.
\subsection{QCar 2 Hardware Validation}

The hardware experiments are conducted using the Quanser QCar 2 platform. The game-theoretic controller is implemented in a MATLAB/Simulink model, deployed through QUARC software, and executed on the onboard NVIDIA Jetson AGX Orin processor. The hardware evaluation includes controller computation, onboard sensing, state estimation, control-command generation, drivetrain actuation, and physical vehicle motion.

The hardware experiments follow the test categories defined in Section~IV-A. Both the ego vehicle and obstacle vehicle begin at $0.0$ m/s unless otherwise specified. Unlike the MATLAB asymmetric-arrival cases, asymmetry in the hardware experiments is introduced through the initial vehicle positions: the ego vehicle begins $1.2$ m from the intersection, and the obstacle vehicle begins $1.0$ m from the intersection.
The hardware tests include the following cases:
\begin{itemize}
\item \textbf{Nash interaction:} Symmetric and asymmetric initial-position cases corresponding to Runs 1A and 1B in Section~IV-A.
\item \textbf{Stackelberg leader-follower interaction:} Symmetric and asymmetric cases corresponding to Runs 2A and 2B in Section~IV-A. In the asymmetric case, the ego vehicle is assigned as leader and begins $1.2$ m from the intersection, while the obstacle vehicle is assigned as follower and begins $1.0$ m from the intersection.
\item \textbf{Hierarchy-conflict interaction:} A leader-leader case and a follower-follower case are evaluated. In both cases, the vehicles begin $1.0$ m from the intersection.
\item \textbf{Robustness interaction:} Constant-speed and unexpected-braking obstacle behaviors are evaluated using the Nash, Stackelberg-leader, and Stackelberg-follower configurations. Both vehicles begin $1.0$ m from the intersection.
\end{itemize}

\begin{table*}[!t]
\caption{MATLAB Simulation Results}
\label{tab:stage1_metrics}
\centering
\begin{tabular}{@{}p{2.0cm} p{2.0cm} c c c p{3cm}@{}}
\toprule
\textbf{Test category} & \textbf{Scenario configuration} & \textbf{Minimum distance (m)} & \textbf{Average speed (m/s)} & \textbf{95th-percentile latency ($\mu$s)} & \textbf{Observed behavior} \\ \midrule
\textbf{Nash} & 1A: Symmetric arrival & $2.804$ & $0.151$ & $2.200$ & Both vehicles yield and remain stationary. \\
 & 1B: Asymmetric arrival & $2.287$ & $0.400$ & $1.500$ & Deadlock resolved by the initial speed difference. \\ \midrule
\textbf{Stackelberg} & 2A: Symmetric leader-follower & $1.262$ & $0.375$ & $6.600$ & The leader proceeds and the follower yields. \\
 & 2B: Asymmetric leader-follower & $1.262$ & $0.199$ & $1.800$ & The assigned hierarchy determines the  actions. \\
& 3: Leader-leader conflict & $0.023$ & $0.375$ & $1.300$ & The obstacle vehicle yields despite the conflicting assignments. \\ \midrule
\textbf{Robustness} & 4A: Constant speed, Nash & $2.287$ & $0.132$ & $1.900$ & The ego vehicle yields to the obstacle vehicle. \\
& 4B: Constant speed, leader & $1.262$ & $0.199$ & $1.500$ & The ego vehicle yields despite being the leader. \\
& 4C: Fast-then-slow, Nash & $2.386$ & $0.124$ & $1.300$ & The controller updates its decision after the obstacle vehicle changes speed. \\
& 4D: Fast-then-slow, follower & $1.268$ & $0.199$ & $2.300$ & The follower yields during the obstacle-vehicle maneuver. \\ \bottomrule
\end{tabular}
\end{table*}

\begin{figure*}[!t]
\centering
\includegraphics[width=\textwidth]{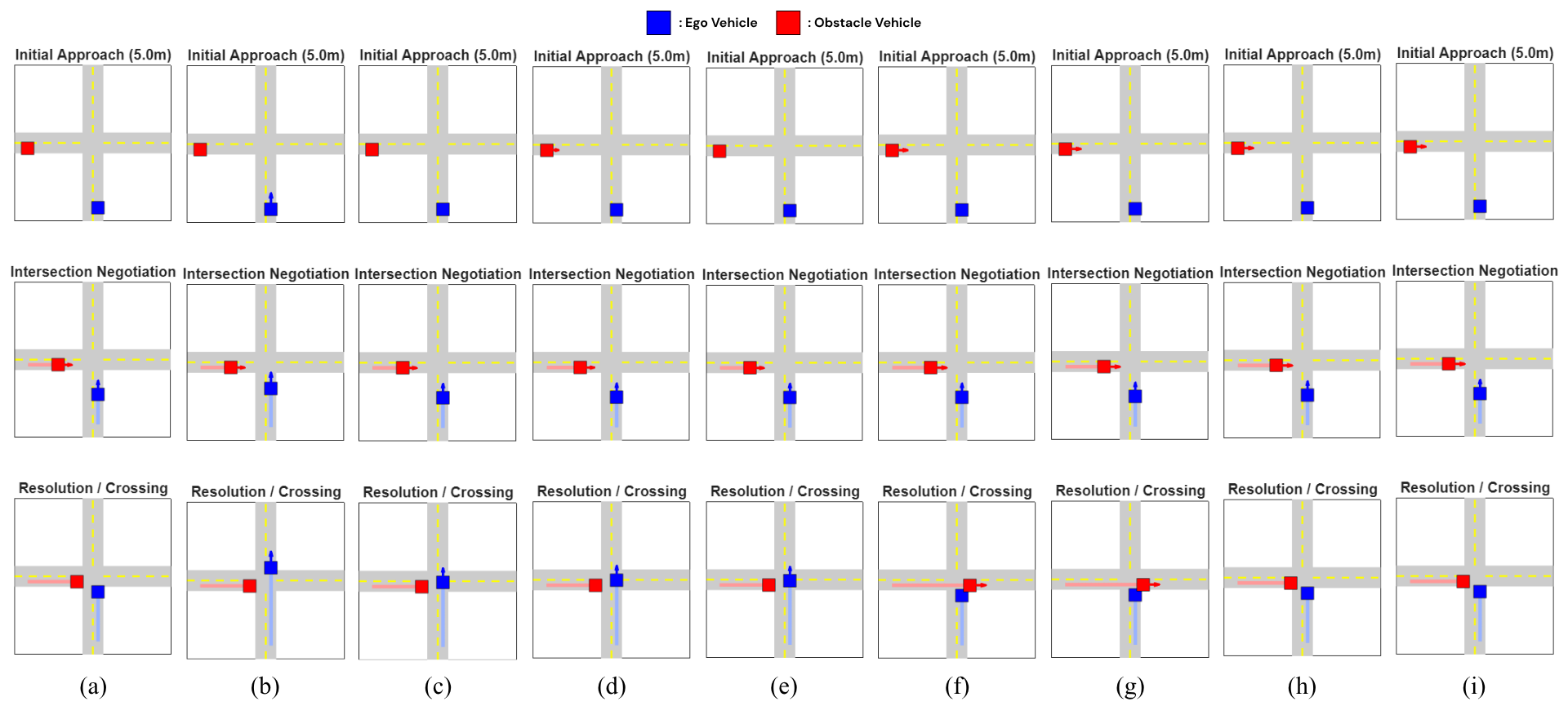}
\caption{MATLAB simulation trajectories for the intersection-crossing cases in Table~\ref{tab:stage1_metrics}. Panels (a)-(i) correspond, from left to right, to the tested Nash, Stackelberg, and robustness cases. Each column shows the initial approach, intersection negotiation, and resolution or crossing state for one scenario.}
\label{fig:stage1_baseline}
\end{figure*}

The controller executes at 100 Hz during the hardware experiments. The reported timing values are obtained from the onboard execution trace.

\begin{table*}[!t]
\caption{QCar 2 Hardware Results}
\label{tab:stage3_metrics}
\centering
\begin{tabular}{@{}p{3.0cm} p{4cm} c c p{4cm}@{}}
\toprule
\textbf{Test category} & \textbf{Scenario configuration} & \textbf{WCET (ms)} & \textbf{Mean latency (ms)} & \textbf{Observed behavior} \\ \midrule
\textbf{Nash} & 1A: Symmetric arrival & $0.1118$ & $0.0366$ & Both vehicles yield and remain stopped. \\
 & 1B: Asymmetric arrival & $0.1273$ & $0.0465$ & The initial position difference produces a passing order. \\ \midrule
\textbf{Stackelberg} & 2A: Symmetric leader-follower & $0.7619$ & $0.2770$ & The leader proceeds and the follower yields. \\
 & 2B: Asymmetric leader-follower & $0.5232$ & $0.2864$ & The selected actions account for hierarchy and relative position. \\
& 3A: Leader-leader conflict & $0.6996$ & $0.2781$ & Conflicting leader assignments result in collision. \\
& 3B: Follower-follower conflict & $0.6965$ & $0.2781$ & Both vehicles yield and avoid collision. \\ \midrule
\textbf{Robustness} & 4A: Constant speed, Nash & $0.2099$ & $0.0391$ & The ego vehicle yields to the obstacle vehicle. \\
& 4B: Constant speed, leader & $0.5840$ & $0.2848$ & The ego vehicle yields when collision cost is high. \\
& 4C: Fast-then-slow, Nash & $0.0925$ & $0.0413$ & The ego vehicle remains stopped after yielding. \\
& 4D: Fast-then-slow, follower & $0.7086$ & $0.2752$ & The follower remains stopped after yielding. \\
& 4E: Fast-then-slow, leader & $0.6540$ & $0.2668$ & The leader selects to proceed after the obstacle decelerates. \\ \bottomrule
\end{tabular}
\end{table*}

\section{Results and Discussion}
\label{sec:results}

\begin{figure*}[!t]
\centering
\includegraphics[width=0.9\textwidth]{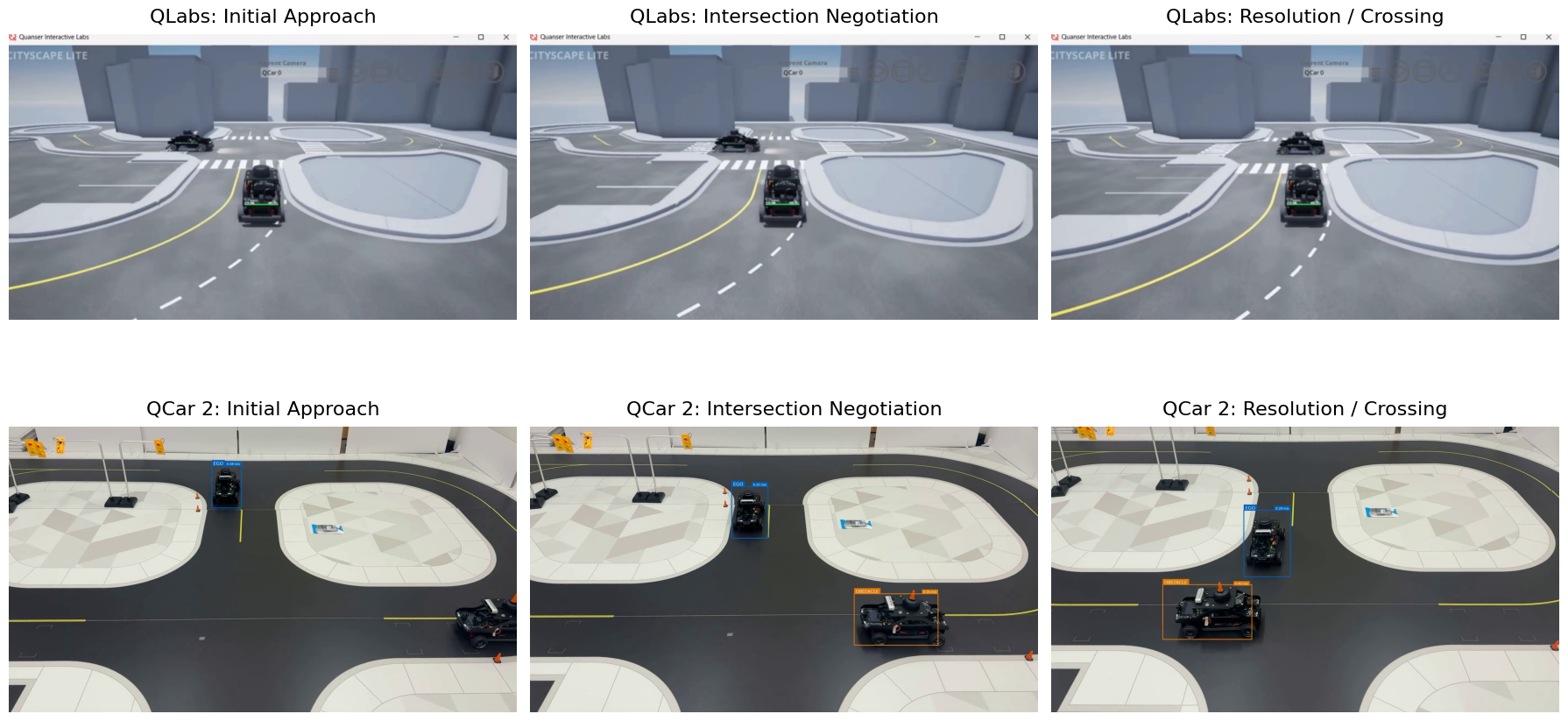}
\caption{Time-lapse progression of the Nash Asymmetric interaction (Run~1B) in the virtual and physical evaluation environments. The top row shows the QLabs simulation, while the bottom row shows the physical QCar~2 hardware experiment. In both environments, the proposed strategy resolves the intersection conflict according to the vehicles' initial kinematic advantage, progressing from approach, through conflict-zone negotiation, to a safe crossing.}
\label{fig:comprehensive_timelapse}
\end{figure*}

\subsection{MATLAB Simulation Results}

The MATLAB results are summarized in Table~\ref{tab:stage1_metrics}. The corresponding trajectories are shown in Fig.~\ref{fig:stage1_baseline}. Panels (a) and (b) show the symmetric and asymmetric Nash cases, respectively. Panels (c) and (d) show the symmetric and asymmetric Stackelberg leader-follower cases, respectively. Panel (e) shows the leader-leader hierarchy conflict, and panels (f)-(i) show the robustness cases. In Table I, the minimum distance refers to the smallest realized Euclidean distance between the centers of mass of the two vehicles, recorded over the full executed trajectory. The average speed denotes the mean longitudinal speed of the ego vehicle, computed across the full duration of the scenario.

In the symmetric Nash case, both vehicles select yielding actions. The resulting behavior avoids collision, but does not resolve the right-of-way conflict. In the asymmetric Nash case, the initial speed breaks the symmetry in the predicted costs, resulting in a passing order. 

The leader-follower Stackelberg cases produce a passing order when the leader and follower roles are consistent across the two vehicles. In the symmetric leader-follower case, the leader proceeds, and the follower yields. In the asymmetric case, the selected action depends on both the hierarchy and the predicted collision cost. The leader assignment does not prevent yielding when the collision penalty becomes dominant.

The leader-leader case illustrates the effect of incompatible hierarchy assumptions. Although both vehicles are assigned leader roles, the obstacle vehicle yields to the ego vehicle in the MATLAB simulation. 

The 95th-percentile MATLAB decision time is below $6.6\,\mu$s for the tested cases. These measurements characterize the desktop MATLAB implementation. Execution times on the embedded platform are evaluated separately. The running time is collected from MATLAB\textsuperscript{\textregistered} on a laptop with an AMD Ryzen 5 220 processor clocked at 3.20~GHz and 16~GB of RAM.

\subsection{Hardware Results}
Table~\ref{tab:stage3_metrics} summarizes the decision times and observed behaviors in the QCar 2 experiments. A video demonstrating the QLabs simulations and physical QCar 2 validation runs across all test cases is available at \textit{\url{https://youtu.be/gkV6lz0twRk}}. For each hardware trace, the computational performance is evaluated using two metrics: the mean latency, which represents the average time required to compute the decision rule during a single control step, and the worst-case execution time (WCET), which records the absolute maximum computation time observed for a control step during the experiment. Both metrics are computed after removing the first $2$ s of data to exclude initialization transients. The maximum recorded WCET is $0.7619$ ms, which is below the $10$ ms period of the 100 Hz control loop used in this work.

The symmetric Nash case produces mutual yielding on the hardware platform. The vehicles remain in a safe halt rather than entering the intersection simultaneously. This result is consistent with the symmetric cost structure, although the physical response includes asynchronous execution, actuator delays, and state-estimation errors that are not represented in the MATLAB model.

In the leader-follower Stackelberg cases, the role assignment provides a passing order. The follower yields when the leader proceeds. In the asymmetric case, the controller also accounts for the relative distance of the vehicles to the intersection. When the predicted collision cost is high, the vehicle assigned as leader can yield instead of entering the conflict region.

Unlike the MATLAB prediction, in which the obstacle vehicle's velocity converges to zero and it yields to the ego vehicle, the leader-leader hardware case results in a collision. In contrast, the follower-follower case results in mutual yielding. These cases show that the Stackelberg implementation requires a consistent hierarchy across the interacting vehicles.

The robustness cases evaluate obstacle vehicles that do not follow the expected game strategy. For constant-speed obstacle vehicles, the ego vehicle yields in both the Nash and Stackelberg configurations. For the unexpected-braking cases, the Nash and Stackelberg-follower configurations remain stopped after the obstacle vehicle decelerates, whereas the Stackelberg-leader configuration selects a passing maneuver after the conflict region becomes available. 

\subsection{Simulation-to-Hardware Observations}

The three-stage evaluation shows that some interaction outcomes are preserved from simulation to hardware, while others are not. In particular, incompatible leader assignments are safely resolved in MATLAB simulation but result in a collision on physical hardware. This result indicates that hierarchy consistency is required for the tested Stackelberg implementation to remain safe under physical deployment.

Other outcomes differ across stages. In the Stackelberg leader crash case, the idealized formulation produces the obstacle leader yielding to the ego leader and allowing safe bypass. On hardware, the leader vehicles did not yield, leading to collision. The software result is safe in the tested experiments, but it does not ensure safety in the hardware experiments. 
This is because the physical controller estimates position and speed using onboard sensing and encoder measurements, which are noisy and are subject to signal discretization inaccuracies that affect the state estimates. The hardware experiments use the same game-theoretic formulation as the MATLAB simulation, but the selected actions depend on the estimated state at each control step.

This discrepancy highlights an important consideration when translating game-theoretic decision-making algorithms to physical systems: robustness to state-estimation errors must be explicitly accounted for in both algorithm design and hardware implementation. Small sensing and discretization errors can alter the perceived interaction state and, consequently, change the equilibrium action selected by the agents. Thus, safety observed under idealized or accurately known states does not necessarily transfer to hardware. Practical implementations should therefore incorporate estimation uncertainty, robustness margins, or safety mechanisms that remain effective when the estimated state deviates from the true physical state.

\section{Conclusion and Future Work}
\label{sec:conclusion}
This paper presented a hardware-in-the-loop evaluation of Nash and Stackelberg game-theoretic controllers for autonomous intersection crossing. The controllers were evaluated through MATLAB simulation, qualitative validation in QLabs, and physical experiments on a Quanser QCar 2 platform. In the tested hardware scenarios, the game-theoretic decision-making computation remained within the 10 ms period of the 100 Hz control loop, with a maximum observed execution time of $0.7619$ ms.
The results indicate that symmetric Nash interactions can lead to mutual yielding and an indefinite halt, whereas consistent Stackelberg leader-follower assignments establish a passing order in the tested scenarios. Conflicting assignments in which both vehicles take the leader role can lead to a collision, while assignments in which both take the follower role can result in mutual yielding. The collision-cost term can cause a leader to yield when proceeding would result in a high predicted collision cost. The hardware results further highlight the importance of accounting for sensing and state-estimation uncertainty when translating game-theoretic controllers from simulation to physical systems.


\section*{Supplementary Material}
A video demonstration of the QLabs simulations and physical QCar~2 hardware experiments is available at \textit{\textbf{\url{https://youtu.be/gkV6lz0twRk}}}.

\bibliographystyle{IEEEtran}
\bibliography{references}

@ARTICLE{liu2022_potential,
  author={Liu, Mushuang and Kolmanovsky, Ilya and Tseng, H. Eric and Huang, Suzhou and Filev, Dimitar and Girard, Anouck},
  journal={IEEE Transactions on Intelligent Transportation Systems}, 
  title={Potential Game-Based Decision-Making for Autonomous Driving}, 
  year={2023},
  volume={24},
  number={8},
  pages={8014-8027},
  doi={10.1109/TITS.2023.3264665},
  ISSN={1558-0016},
  month={Aug},}

@article{bateman2023,
title = {Nash or Stackelberg? – A Comparative Study for Game-Theoretic Autonomous Vehicle Decision-Making},
journal = {IFAC-PapersOnLine},
volume = {58},
number = {28},
pages = {504-509},
year = {2024},
note = {The 4th Modeling, Estimation, and Control Conference – 2024},
issn = {2405-8963},
doi = {https://doi.org/10.1016/j.ifacol.2025.01.096},
url = {https://www.sciencedirect.com/science/article/pii/S2405896325000965},
author = {Brady Bateman and Ming Xin and H. Eric Tseng and Mushuang Liu}
}

@ARTICLE{liu2022_predictor,
  author={Liu, Mushuang and Tseng, H. Eric and Filev, Dimitar and Girard, Anouck and Kolmanovsky, Ilya},
  journal={IEEE Transactions on Control Systems Technology}, 
  title={Safe and Human-Like Autonomous Driving: A Predictor–Corrector Potential Game Approach}, 
  year={2024},
  volume={32},
  number={3},
  pages={834-848},
  doi={10.1109/TCST.2023.3332438},
  ISSN={1558-0865},
  month={May},}

@article{liu2023_projection,
author = {Liu, Mushuang and E. Tseng, H. and Filev, Dimitar and Girard, Anouck and Kolmanovsky, Ilya},
title = {Game Projection and Robustness for Game-Theoretic Autonomous Driving},
year = {2025},
issue_date = {March 2025},
publisher = {IEEE Press},
volume = {26},
number = {3},
issn = {1524-9050},
url = {https://doi.org/10.1109/TITS.2024.3514826},
doi = {10.1109/TITS.2024.3514826},
journal = {IEEE Transactions on Intelligent Transportation Systems},
month = mar,
pages = {3446–3457},
numpages = {12}
}

@ARTICLE{li2019,
  author={Li, Nan and Yao, Yu and Kolmanovsky, Ilya and Atkins, Ella and Girard, Anouck R.},
  journal={IEEE Transactions on Intelligent Transportation Systems}, 
  title={Game-Theoretic Modeling of Multi-Vehicle Interactions at Uncontrolled Intersections}, 
  year={2022},
  volume={23},
  number={2},
  pages={1428-1442},
  doi={10.1109/TITS.2020.3026160},
  ISSN={1558-0016},
  month={Feb},}

@ARTICLE{hang2020,
  author={Hang, Peng and Lv, Chen and Xing, Yang and Huang, Chao and Hu, Zhongxu},
  journal={IEEE Transactions on Intelligent Transportation Systems}, 
  title={Human-Like Decision Making for Autonomous Driving: A Noncooperative Game Theoretic Approach}, 
  year={2021},
  volume={22},
  number={4},
  pages={2076-2087},
  doi={10.1109/TITS.2020.3036984},
  ISSN={1558-0016},
  month={April},}

@ARTICLE{fabiani2020,
  author={Fabiani, Filippo and Grammatico, Sergio},
  journal={IEEE Transactions on Intelligent Transportation Systems}, 
  title={Multi-Vehicle Automated Driving as a Generalized Mixed-Integer Potential Game}, 
  year={2020},
  volume={21},
  number={3},
  pages={1064-1073},
  doi={10.1109/TITS.2019.2901505},
  ISSN={1558-0016},
  month={March},}

@ARTICLE{tian2019,
  author={Tian, Ran and Li, Nan and Kolmanovsky, Ilya and Yildiz, Yildiray and Girard, Anouck R.},
  journal={IEEE Transactions on Intelligent Transportation Systems}, 
  title={Game-Theoretic Modeling of Traffic in Unsignalized Intersection Network for Autonomous Vehicle Control Verification and Validation}, 
  year={2022},
  volume={23},
  number={3},
  pages={2211-2226},
  doi={10.1109/TITS.2020.3035363},
  ISSN={1558-0016},
  month={March},}

@misc{yan2026,
      title={Markov Potential Game and Multi-Agent Reinforcement Learning for Autonomous Driving}, 
      author={Huiwen Yan and Mushuang Liu},
      year={2026},
      eprint={2603.19188},
      archivePrefix={arXiv},
      primaryClass={eess.SY},
      url={https://arxiv.org/abs/2603.19188}, 
}

@ARTICLE{liu2020_computing,
  author={Liu, Liangkai and Lu, Sidi and Zhong, Ren and Wu, Baofu and Yao, Yongtao and Zhang, Qingyang and Shi, Weisong},
  journal={IEEE Internet of Things Journal}, 
  title={Computing Systems for Autonomous Driving: State of the Art and Challenges}, 
  year={2021},
  volume={8},
  number={8},
  pages={6469-6486},
  doi={10.1109/JIOT.2020.3043716},
  ISSN={2327-4662},
  month={April},}

@misc{geary2020,
      title={Resolving Conflict in Decision-Making for Autonomous Driving}, 
      author={Jack Geary and Subramanian Ramamoorthy and Henry Gouk},
      year={2021},
      eprint={2009.06394},
      archivePrefix={arXiv},
      primaryClass={cs.GT},
      url={https://arxiv.org/abs/2009.06394}, 
}

@misc{liu2026_time,
      title={Real-Time Solution-Seeking for Game-Theoretic Autonomous Driving via Time-Distributed Iterations}, 
      author={Shaoqing Liu and Mushuang Liu},
      year={2026},
      eprint={2604.16184},
      archivePrefix={arXiv},
      primaryClass={eess.SY},
      url={https://arxiv.org/abs/2604.16184}, 
}

@ARTICLE{namazi2019,
  author={Namazi, Elnaz and Li, Jingyue and Lu, Chaoru},
  journal={IEEE Access}, 
  title={Intelligent Intersection Management Systems Considering Autonomous Vehicles: A Systematic Literature Review}, 
  year={2019},
  volume={7},
  number={},
  pages={91946-91965},
  doi={10.1109/ACCESS.2019.2927412},
  ISSN={2169-3536},
  month={},}

@techreport{muhlethaler2026,
  TITLE = {{Automated Intersection Management for Connected and Automated Vehicles: A Twenty-Five-Year Review}},
  AUTHOR = {M{\"u}hlethaler, Paul},
  URL = {https://hal.science/hal-05648765},
  INSTITUTION = {{AIO - Inria Paris}},
  YEAR = {2026},
  MONTH = Jun,
  HAL_ID = {hal-05648765},
  HAL_VERSION = {v1},
}

@INPROCEEDINGS{suriyarachchi2022,
  author={Suriyarachchi, Nilesh and Chandra, Rohan and Baras, John S. and Manocha, Dinesh},
  booktitle={2022 IEEE 25th International Conference on Intelligent Transportation Systems (ITSC)}, 
  title={GAMEOPT: Optimal Real-time Multi-Agent Planning and Control for Dynamic Intersections}, 
  year={2022},
  volume={},
  number={},
  pages={2599-2606},
  doi={10.1109/ITSC55140.2022.9921968},
  ISSN={},
  month={Oct},}

@ARTICLE{hu2023,
  author={Hu, Xuemin and Li, Shen and Huang, Tingyu and Tang, Bo and Huai, Rouxing and Chen, Long},
  journal={IEEE Transactions on Intelligent Vehicles}, 
  title={How Simulation Helps Autonomous Driving: A Survey of Sim2real, Digital Twins, and Parallel Intelligence}, 
  year={2024},
  volume={9},
  number={1},
  pages={593-612},
  doi={10.1109/TIV.2023.3312777},
  ISSN={2379-8904},
  month={Jan},}

@ARTICLE{stocco2021,
  author={Stocco, Andrea and Pulfer, Brian and Tonella, Paolo},
  journal={IEEE Transactions on Software Engineering}, 
  title={Mind the Gap! A Study on the Transferability of Virtual Versus Physical-World Testing of Autonomous Driving Systems}, 
  year={2023},
  volume={49},
  number={4},
  pages={1928-1940},
  doi={10.1109/TSE.2022.3202311},
  ISSN={1939-3520},
  month={April},}

@INPROCEEDINGS{lambertenghi2025,
  author={Lambertenghi, Stefano Carlo and Flores Valdez, Mirena and Stocco, Andrea},
  booktitle={2025 40th IEEE/ACM International Conference on Automated Software Engineering (ASE)}, 
  title={A Multi-Modality Evaluation of the Reality Gap in Autonomous Driving Systems}, 
  year={2025},
  volume={},
  number={},
  pages={2808-2820},
  doi={10.1109/ASE63991.2025.00230},
  ISSN={2643-1572},
  month={Nov},}

@misc{quanser_qcar2,
  title        = {{QCar 2: 1/10th Scale Autonomous Vehicle Platform Information Sheet}},
  author       = {{Quanser Inc.}},
  organization = {Quanser Inc.},
  address      = {Markham, ON, Canada},
  year         = {2024},
  note         = {[Online]. Available: \url{https://www.quanser.com/products/qcar-2/}}
}

\end{document}